# The birth of a quasiparticle in Si observed in time-frequency space

Muneaki Hase\*, Masahiro Kitajima\*, Anca Monia Constantinescu† & Hrvoje Petek†

*\*Materials Engineering Laboratory, National Institute for Materials Science,*

*1-2-1 Sengen, Tsukuba, Ibaraki 305-0047, Japan*

*†Department of Physics and Astronomy, University of Pittsburgh, Pittsburgh, PA 15260, USA*

**Keywords:** *silicon, quasiparticle, electron-phonon interaction, ultrafast spectroscopy.*

The concept of quasiparticles in solid-state physics is an extremely powerful way to describe complex many-body phenomena in terms of single particle excitations.[1] Introducing a simple particle such as electron, *e*, hole, *h*, or a phonon, *p*, deforms a many-body system through interaction with other particles. We say the added particle is "dressed" or "renormalized" by a self-energy cloud that describes the response of the many-body system forming a new entity, the quasiparticle. With ultrafast laser techniques we can impulsively generate bare particles and observe their dressing by the many-body interactions, that is quasiparticle formation, on the time and energy scales governed by the Heisenberg uncertainty principle.[2] Here we present the coherent response of Si to excitation with a 10 femtosecond ($10^{-14}$ s) laser pulse. The optical pulse interacts with the sample via the complex second-order nonlinear susceptibility to generate a force on the lattice driving coherent phonon excitation. Transforming the transient reflectivity signal into frequency-time space reveals interference effects leading to the coherent phonon generation and subsequent dressing of the phonon by electron-hole, *e-h*, pair excitations.

Silicon is a paradigm for the fundamental electronic and mechanical properties of Group IV and III-V semiconductors and the basic material for the electronic device technology. The interaction between electrons and phonons via the Coulomb force defines many of the physical properties of Si, for instance, the electrical conductivity: real and virtual scattering processes transfer carriers between different momentum states and renormalize their mass.[1] Much of our knowledge of the electron-phonon interaction in Si is limited to transport measurements at near equilibrium conditions; however, as the electronic device dimensions become comparable to carrier scattering lengths, the fundamental quantum mechanical carrier-carrier and carrier-lattice interactions increasingly determine their properties.[3] Measurements of the carrier relaxation in silicon by optical techniques have been restricted mainly to transient changes of index of refraction following excitation by intense laser pulses,[4-7] because the indirect band gap of Si obviates more specific probes. It has been established that the quantum mechanical phase of the primary photo-excited *e-h* pairs (momentum scattering) decays in «100 fs, while charge carriers equilibrate with the lattice on 200 – 300 fs time scale.[5-9]

Here we explore the coherent response of the coupled Si carrier-lattice system to excitation by a 10 fs laser pulse. The transient anisotropic reflectivity signal is measured and transformed into time-frequency space revealing the dynamics of formation of the zone center (zero momentum) coherent longitudinal optical (LO) phonon at 15.3 THz,[6] and its subsequent dressing by interaction with the photogenerated carrier plasma. The birth of a quasiparticle, the coherent phonon, is observed in the quantum kinetic regime where energy-time uncertainty and phase memory stored by the coherent polarization make the carrier-light, carrier-carrier, and carrier-phonon interactions non-local in time.[2]

The anisotropic transient reflectivity of n-doped Si(001) (doping concentration $\sim 1 \times 10^{15}$ cm$^{-3}$) under ambient conditions is measured employing the electro-optic sampling method.[10] Nearly collinear, pump and probe pulses (10 fs pulse duration; 406 nm (3.05 eV) wavelength; 80 MHz repetition rate) are focused to a 10 μm spot on the sample with an average angle of 5° from the surface normal. The pump pulse with 40 mW average power generates a carrier density of $4 \times 10^{19}$ cm$^{-3}$; the probe power is $\leq 2$ mW. Polarization of the two beams is linear with a mutual angle of 45°, as shown in inserts of Figure 1. After reflecting from the sample, the probe beam is analyzed into polarization components parallel and perpendicular to that of the pump, $R_\parallel$ and $R_\perp$, and detected with photodiodes. The resulting photocurrents are subtracted, and their difference $\Delta R_{eo}(\tau) = R_\perp(\tau) - R_\parallel(\tau)$ recorded as a function of the pump-probe delay $\tau$.

The excitation light interacting with the sample generates nonlinear polarization $P_j^2(\tau)$ normal to the surface by the process of optical rectification,

$$P_j^2(\tau) = \chi_{jkl}^2(\tau - t) E_k(t) E_l(t) + \int_{-\infty}^{\tau} J_j(t') dt', \qquad (1)$$

where $\chi_{jkl}^2(\tau - t)$ is the complex, second-order nonlinear susceptibility, $E_{k,l}(t)$ are vectorial components of the incident electric fields, and $J_j(t')$ is the macroscopic photo-Dember current arising from differential electron/hole mobilities.[11-13] Since the bulk dipole contribution to $\chi_{jkl}^2(\tau - t)$ is zero by symmetry, the bulk quadrupole and surface dipole terms probably dominate the electro-optic response.[14] The optical rectification or inverse electro-optic response described by $\chi_{jkl}^2\{0;-\omega,\omega'\}$ generates coherent polarization at the beat frequency $(\omega - \omega' \approx 0)$ by the difference frequency mixing of components of the carrier wave $(\omega,\omega')$.[11-13] Because under resonant conditions the optical rectification involves *e-h* pair excitation with finite phase and energy relaxation times, the $\chi_{jkl}^2(\tau - t)$ response depends on the delay $\tau$. The surface-normal [001] component of the induced nonlinear polarization modulates the sample reflectivity via the electro-optic effect,[10]

$$\frac{R_\perp(\tau) - R_\parallel(\tau)}{R_0} = \frac{\Delta R_{eo}(\tau)}{R_0} \approx rP_{001}^2(\tau), \qquad (2)$$

where the coefficient $r$ represents the bulk quadrupole and surface dipole electro-optic response. Because in the electro-optic sampling mode the longitudinal polarization $P_{001}^2(\tau)$ modulates $R_\perp$ and $R_\parallel$ with the opposite phase, the difference $\Delta R_{eo}(\tau)$ records only the dynamics of $P_{001}^2(\tau)$, while the much larger signal from the isotropic change in reflectivity due to the incoherent carrier and lattice dynamics is eliminated.[10] The cascaded sequence of excitation and probing can be described by third-order susceptibility tensor $\chi^3(\omega'';\omega',-\omega,\omega') \approx \chi^2(0;-\omega,\omega') \cdot \chi^2(\omega'';\omega',0)$ corresponding to the coherent Stokes and anti-Stokes Raman scattering (CSRS and CARS; Figure 2a), as long as the coherent polarization exists in the interval between the pulses.[15] Other fully resonant intrinsic $\chi^3$ processes, such as the induced birefringence, which involve similar excitations to the cascaded processes, can also contribute to the signal.

Transient reflectivity is measured both with the [110] crystalline axis of Si oriented parallel and at 45° to the pump polarization to explore the $\Gamma_{25'}$ and $\Gamma_{12}$ symmetry response, respectively (see Figure 1). The $\Delta R_{eo}(\tau)/R_0$ measurement for the $\Gamma_{25'}$ geometry in Figure 1a is composed of an aperiodic component near zero delay followed by a coherent LO phonon oscillation. The latter component can be fit by $A_0 \exp(-\tau/1.30)\cos\left[(15.24*\tau + 0.016\tau^2 + 0.064)*2\pi\right]$, where the parameters in the order of appearance are the dephasing time in picoseconds, the oscillation frequency in THz, chirp, and phase shift. The departure from the LO phonon dephasing time of 3.5 ps and frequency of 15.60 THz of intrinsic Si reflects the coherent phonon self-energy, which depends on the excitation density.[16] By contrast, the $\Gamma_{12}$ response in Figure 1b consists of a half-cycle transient reminiscent of the aperiodic component in the $\Gamma_{25'}$ response.

To glean more insight into the early time dynamics, Figures 1c and d show the $\Delta R_{eo}(\tau)/R_0$ signals converted into time-dependent spectral amplitudes (chronograms) by continuous wavelet transform (CWT).[17] The CWT decomposes the signal into an ultra-broadband (DC to >70 THz) response straddling zero delay in both geometries, and the LO phonon response at ~15.3 THz present only in the $\Gamma_{25'}$ geometry. The fast response is the coherent electronic coupling of the pump and probe fields via the nonlinear susceptibility of the sample commonly referred to as the coherent artifact. This misnomer belies that in absorbing matter the laser induced coherent charge density fluctuations include specific microscopic electronic excitation and scattering processes with finite dephasing times.[18] The real and virtual processes associated with near-resonant photoexcitation across the direct band gap of Si, which contribute to $\chi^2$ and $\chi^3$, both modulate the anisotropic reflectivity and generate the driving force for the coherent phonon excitation.[19] The data in Figure 1a and c resolve the buildup of the electronic force and the ensuing response of the lattice that constitute the electron-phonon interaction in solid-state materials.

Processes contributing to $\chi^2$ and $\chi^3$ are photon absorption, and resonant electronic and vibrational Raman scattering. According to Figure 2b, 3.05 eV light is near resonant with the $E_0'$ and the more intense $E_1$ critical points at 3.320 eV and 3.396 eV, which correspond to the excitation *e-h* pairs with near-zero momentum ($\Gamma$ point), and for a range of momenta between $(2\pi/a)\langle\frac{1}{4},\frac{1}{4},\frac{1}{4}\rangle$ and $(2\pi/a)\langle\frac{1}{2},\frac{1}{2},\frac{1}{2}\rangle$ along the $\Lambda$ line, respectively.[20] Moreover, for the alignment of the pump polarization in the $\Gamma_{25'}$ geometry (Figure 1 c insert) only two out-of-four tetrahedral bonds of each Si atom are excited.[18,21] This real space anisotropy translates to the excitation of electrons to 4 out-of 8 equivalent local minima (L-valleys) and holes along $\Lambda$ lines in the reciprocal space. This anisotropic distribution exerts a force on the lattice through an imbalance of deformation potentials, which describe distortion of the lattice in response to charge density fluctuations,[18,19] and of chemical potentials between different nonthermal carrier populations.[21] These two carrier-lattice interactions, which contribute to the imaginary part of the Raman tensor, exert a step function electrostrictive force on the lattice driving the coherent LO phonon oscillation polarized in the [001] direction.[19] By contrast, equivalent excitation of each bond in the $\Gamma_{12}$ geometry balances the forces to zero precluding the LO phonon excitation (Figure 1d inset).

In addition to absorption, electronic Raman scattering contributes microscopic currents to the longitudinal polarization.[11,16,21-23] The anisotropic inelastic intraband scattering of light by free-electrons, which is unscreened, occurs in materials with non-spherical Fermi surfaces, e.g. the valence and conduction bands of Si.[24,25] In addition to intraband scattering, the inter-valence band electronic Raman scattering among the heavy-hole, light-hole, and split-off bands indicated in Figure 2c also contribute significantly to the $\chi^2$ response predominantly in the $\Gamma_{25'}$ geometry.[26] These electronic contributions have distinct spectral signatures for thermal carriers in n- and p-type Si.[16,22,25] Intraband scattering measures the carrier velocity distribution with the energy transfer given by $\Delta\omega \approx qV$, where $q$ is the scattering wave vector and $V$ the carrier velocity.[24,25] The intraband spectrum forms a Lorentzian or Gaussian wing on the elastic scattering peak with a carrier density and temperature dependent width of typically 6 THz.[22] When calculated with realistic band structure of Si, the interband component, has a single broad maximum at ~20 THz.[26,27] Because it overlaps with the intraband component, the interband component can be identified clearly only for hole doping of >$10^{20}$ cm$^{-3}$.[23] Since the photogenerated carrier density is significantly higher than the doping density, the electronic Raman contributions are mainly from the photogenerated non-thermal plasma. Therefore, electronic spectral distributions are modified from the thermal ones, and their contribution rises with the photogenerated plasma density.

To identify different contributions to the coherent response of Si that appear in the CWT chronograms, Figure 3 presents cross sections of the $\Gamma_{25'}$ and $\Gamma_{12}$ chronograms at zero delay for 5 and 40 mW excitation. Three distinct features "A", "B", and "C" can be identified. The component "A" is assigned to the intraband electronic scattering based on its expected frequency. The calculated interband Raman spectra are qualitatively similar to the main component of the electronic response with the maximum at ~13 THz ("C" in Figure 3).[26] The width of this component broadens for high pump-power particularly in the $\Gamma_{12}$ geometry indicating that it depends on the carrier density and distribution. However, the interband scattering cannot make exclusive contribution to "C", because the response is nearly identical in the $\Gamma_{12}$ and $\Gamma_{25'}$ geometries, while the interband scattering is mainly observable in the $\Gamma_{25'}$ geometry. We attribute the "C" component to all electronic processes that contribute to the $\chi^3$ response including the *e-h* pair generation (photoabsorption), and the electronic Raman scattering from the hot plasma.

However, the most intriguing aspect of the coherent response of Si is the anti-resonance (dip) at 15.3 THz in the $\Gamma_{25'}$ geometry ("B" in Figure 3). The anti-resonance in CWT chronograms has the maximum negative amplitude in the region of overlap between the coherent phonon and electronic responses at 22 fs and 15.3 THz, suggesting that it is related to the coherent excitation of the coupled phonon-carrier systems. The precise location of the anti-resonance in time and frequency depends on the doping of the sample and the pump power indicating that it is sensitive to the type of dopant and the distribution of nonthermal carriers. The evolution of the anti-resonance such as its delayed appearance, width that narrows with delay, and position that shifts to higher frequency are suggestive of quantum mechanical interference phenomena that arise from the energy-time uncertainty, parallel excitation pathways, and finite phase relaxation time of the induced polarization (~16 fs from the Lorentzian linewidth of 80 meV at 300 K for the $E_1$ critical point).[8,21,28]

We attribute the anti-resonance in the CWT chronograms of Si to the coupling of the LO phonon and *e-h* pair continuum amplitudes via the $\chi^3$ scattering process and electron-phonon interaction. Such discrete-continuum coupling is generic to many physical systems and is manifest in the well known Fano spectral lines shapes.[29] The anisotropic reflectivity measurement in Figure 1a captures the many-body interactions involving the valance band *e-h* pair and LO phonon excitations that gives rise to the Fano line shapes in the LO phonon Raman spectra of highly *p*-doped Si.[16,23] By projecting the data into frequency-time space we can follow the interactions that give birth to coherent phonons.[16,27,29,30] The $\Gamma_{25'}$ chronograms represent the joint electronic and phonon scattering probabilities coherently coupled via the second-order Raman transition moment and the electron-phonon interaction. Green's function derivations of discrete-continuum spectral functions attribute anti-resonances to final-state interactions and asymmetric line shapes to interference effects, both stemming from the fundamental interaction that allows phonons to decay into and be generated from the *e-h* pair continuum.[27,29,30] The anti-resonance is a manifestation of final state interactions through which the LO phonon amplitude is dissolved in the *e-h* pair continuum and the *e-h* pair continuum amplitude is suppressed at the resonance frequency. The asymmetry of line shapes arises from the interference between spectral amplitudes, which has the origin in the *−arctan* phase relationship between the electronic and lattice responses.[29] The electronic force leads the LO phonon response by $\pi/2$ at resonance; on either side of the resonance, constructive and destructive interference occurs for $<\pi/2$ and $>\pi/2$, respectively. The distinct coherent LO phonon amplitude emerges only after ~100 fs delay when the electronic coherences have substantially dephased. The residual manifestation electron-phonon interaction is the carrier density-dependent LO phonon frequency and decay rate representing the real and imaginary parts of self-energy.

The CWT chronograms in Figure 1 provide a spectacular frequency-time survey of the discrete-continuum interaction, a pervasive feature in physics. The transient reflectivity measurements demonstrate the possibility of observing in real time the quantum mechanical manifestations of carrier-phonon interactions in Si, which so far could only have been deduced from transport measurements and spectral lineshape analysis. The observation of the coherent response to impressed external fields at >50 THz frequencies defines the fundamental response of silicon, and it opens the way to application and manipulation of internal fields and forces that could lead to applications in ultra-broadband signal processing at frequencies that are intrinsically orders-of-magnitude faster than conventional electronics.

**Acknowledgements.** The authors thank D. Boyanovsky, A. P. Heberle, and K. Ishioka for illuminating discussions. This research has been supported by NSF, the University of Pittsburgh, Grant-in-Aid for Scientific Research from MEXT of Japan, and NIMS Research Funds.

**Competing interests statement.** The authors declare to have no competing financial interests.

**Supplementary material.** Three-dimensional CWT animations can be viewed at http://www.phyast.pitt.edu/People/Faculty/H_Petek/3Dwavelet.

Correspondence and requests for materials should be addressed to H. P. (petek@pitt.edu).

**Figure Captions**

**Figure 1.** Transient electro-optic reflectivity signal for Si(001) in a) $\Gamma_{25'}$ and b) $\Gamma_{12}$ geometry. c) and d) are the continuous wavelet transforms (chronograms) of the data in a) and b), respectively. Left inserts in c) and d) define the polarization of the laser beams relative to the crystalline axes of the sample in the $\Gamma_{25'}$ and $\Gamma_{12}$ geometries. Blue spheres indicate Si atoms in the diamond lattice with size denoting upward or downward disposition with respect to the central atom. Excited or quiescent bonds are represented by red and gray. The signal in a) and b) is derived from the difference in $R_{\parallel}$ and $R_{\perp}$, the orthogonal polarization components of the probe beam. The color scale indicates the signal amplitude red being the maximum.

**Figure 2.** a) The excitation scheme for electro-optic sampling. The pump pulse generates coherent polarization $P_{001}^2$ at a frequency $\omega - \omega'$. The delayed probe beam scatters from the polarization generating light at $2\omega - \omega'$ frequency. The cascaded interaction corresponds to coherent anti-Stokes Raman scattering (CARS). b) The band structure of Si for momenta in the $\Lambda$ and $\Delta$ directions. Photoabsorption at 3.05 eV occurs near the $E_0'$ and $E_1$ critical points, corresponding to transitions near the Brillouin zone center ($\Gamma$ point) and a range of momenta along the $\Lambda$ line, indicated by blue arrows and shading. c) Electronic Raman processes that can contribute to the nonlinear polarization. Intraband scattering occurs mainly within the heavy-hole (HH) valence band, and the X- and L-valleys (blue arrows) in the conduction band. Interband scattering involves transitions among the heavy-hole (HH), light-hole (LH) and split-off (SO) bands (SO→HH excitation is indicated by red arrows). Interaction between the SO→HH continuum and the LO phonon (excitation energy indicated by green shading) contributes to the renormalization of the phonon frequency. The yellow shading conveys the relaxation of electrons towards conduction band minima and holes towards the valence band maximum.

**Figure 3.** Slices of continuous wavelet transforms (CWT) in Figure 1 at zero delay for pump powers of 40 mW (top) and 5 mW (bottom). The cuts show different components A, B, and C (explained in the text) that contribute to the nonlinear polarization. The amplitude suppression at zero frequency is an artifact of the CWT procedure and finite measurement time.

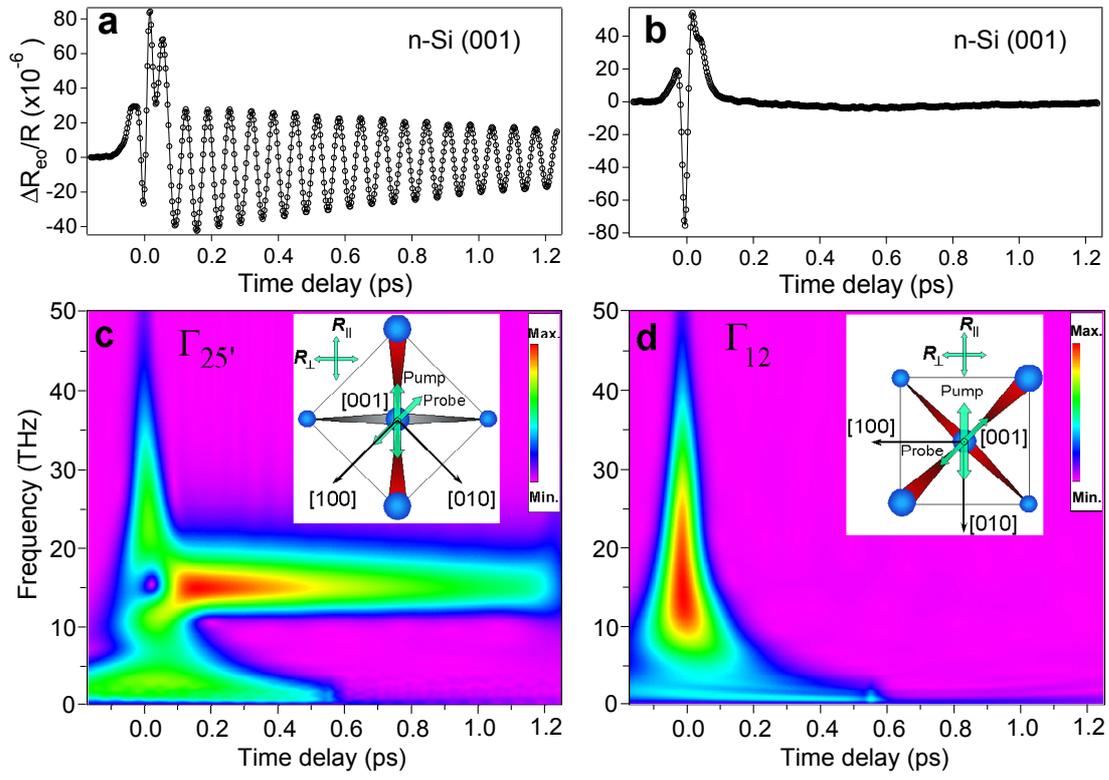

Fig.1. Hase *et al.*

Fig. 2, Hase *et al.*

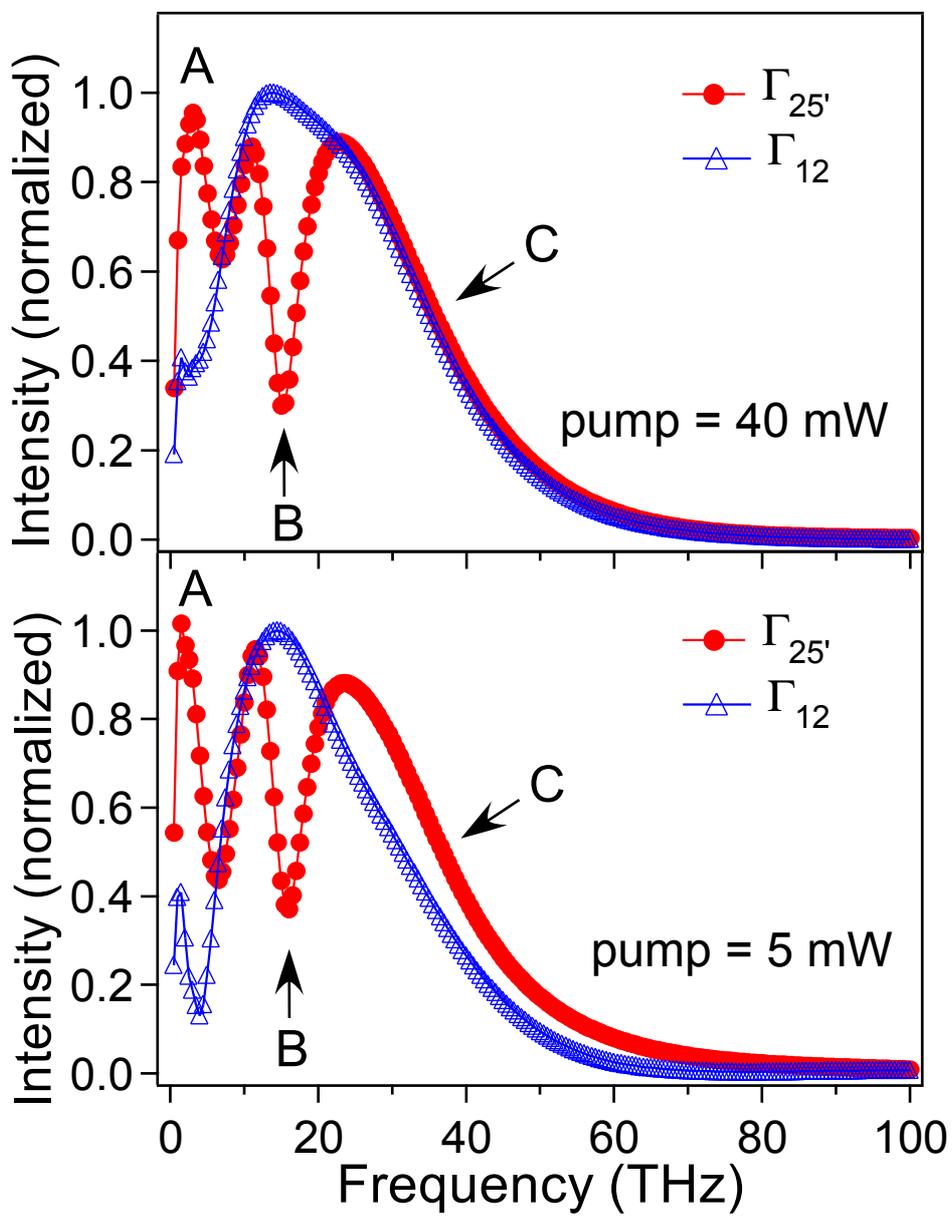

Fig. 3. Hase *et al.*